\documentstyle[prd,aps,floats,epsf]{revtex}
\newcommand{\be}{\begin{equation}}
\newcommand{\ee}{\end{equation}}
\newcommand{\bea}{\begin{eqnarray}}
\newcommand{\eea}{\end{eqnarray}}
\newcommand{\bd}{\begin{displaymath}}
\newcommand{\ed}{\end{displaymath}}

\begin{document}
\bibliographystyle{physics}
\renewcommand{\thefootnote}{\fnsymbol{footnote}}

\author{
Hongying Jin$^1$ and Xinmin Zhang$^2$
\thanks{Email: jhy@zimp.zju.edu.cn,~~xmzhang@mail.ihep.ac.cn}\\
{\small\sl  $^1$Institute of Modern physics, Zhejiang
University}\\
{\small\sl $^2$ Institute of High Energy Physics, Academia
Sinica, P.O.Box 918(4), Beijing 100039, China }\\
}

\title{
\vspace{3cm}
\bigskip
\bigskip
{\LARGE\sf Scalar Glueball Decay Into Pions In Effective Theory}
 }
\maketitle
\vspace{3cm}
\begin{abstract}

We discuss the mixing between the sigma meson $\sigma$
and the "pure" glueball field $H$ and
study the decays of the scalar glueball
candidates $f_0(1370)$, $f_0(1500)$ and $f_0(1710)$ (a linear
combination of
$\sigma$ and $H$) into two pions in an effective linear sigma model.

\end{abstract}

\vspace{1.5cm}
{\bf PACS  numbers 13.25.Hw 12.28.Lg }
\newpage

Quantum Chromo Dynamics (QCD), as
the fundamental theory of the strong
interaction predicts the existence of exotic mesons
made of gluons. Observations
of these gluonium will provide a
direct confirmation on the special feature of non-abelian
gauge theory.\\

Lattice QCD\cite{latt} and various phenomenological models,
such as the potential models\cite{bag} flux tube\cite{flux} all
predict the lowest-lying glueball is $0^{++}$ state, whose mass
is around 1.6GeV.
However it is very difficult to identify glueballs experimentally and
distinquish them from the normal $\bar qq$ states.
So far  five isoscalar states, $\sigma(400-1200)$,
$f_0(980)$, $f_0(1370)$, $f_0(1500)$, $f_0(1710)$ confirmed
experimentally as $0^{++}$ states in the region
1000GeV-1700GeV\cite{data}. It is beyond the accomodation of the naive
quark model if there is only one nonet in this energy region.
The excited states are generally expected to be much heavier. Recently,
some new approaches have been proposed, which tried to include these
states into two nonets\cite{tor1}\cite{pen3}. These models should be
confirmed further by the  experiemnts. Now let us consider these states
in more detail.

The identification of the sigma particle $\sigma$ is
highly controversial.
It is listed in the Particle Data Tables\cite{data} as a very broad meson
with mass
around $400 - 1200$ MeV and full width $600-1200$ MeV. This particle,
needed in the linear sigma model, has been argued to play an important
role
in nuclear physics\cite{ta} and in the study on the chiral phase
transition\cite{pisarski}.
 Recently,
experiments on $D\rightarrow
3\pi$ in E791\cite{E791a} and $J/\psi\rightarrow\omega\pi\pi$ in
BES\cite{BES1a} provide strong evidences on the existence of
$\sigma(400-1200)$. And the theoretical studies on
these processes\cite{E791b}
\cite{BES1b}show that the linear $\sigma$ model gives rise to a reasonable
description of the $\sigma$ decay into $\pi$'s.

It should be pointed out that the authors of Refs.
\cite{kiss}\cite{Och}take
$\sigma(400-1200)$ as a glueball,
because
$\sigma(400-1200)$ does not appear in
$\gamma\gamma\rightarrow\pi\pi$. Pennington later pointed out
that the gauge invariance requires a mass-dependent fit in
$\gamma\gamma\rightarrow\pi\pi$\cite{Pen}, then the appearance of
$\sigma(400-600)$ is still possible. T\"ornqvist used linear sigma model
to
give a very good fit to the light meson spectrum\cite{torn}. The most
recently,  Pennington {\it et. al.} showed that $\sigma$ has a large
content of $\bar uu+\bar dd$ by the finite energy QCD sum rule
approach\cite{Pen1}.
Therefore, in this paper we take $\sigma(400-1200)$ as a lightest
$0^{++}$ $\bar qq$ meson.  $f_0(980)$ is often interpreted as a
multiquark state,{\it i.e.} $\bar KK$ bound state or $\bar qq$\cite{data},
because it strongly couples to $\bar KK$. $f_0(1370)$ has a very broad
decay width and is generally considered as a $\bar qq$  state,
however, a recent fit indicated that it has a large gluon content\cite{close1}.
Therefore, we will take it as a glueball candidate in this
paper.  $f_0(1500)$ and $f_0(1710)$ are the most possible
candidates for the glueballs up to now, not only because their masses are
very close to the predictions of Lattice QCD, but also because they have
very narrow widths and couple to photons weakly. Besides, $f_0(1500)$ also
has enhanced production rate at low transverse momentum
transfer in central collisions\cite{data}.
 Regarding the role of the sigma particle in the
glueball physics,
BES\cite{BES} and MKIII\cite{MKIII} data on
$B(J/\Psi\rightarrow\gamma f_0(1500)\rightarrow
\gamma 4\pi) $ and
$B(J/\Psi\rightarrow\gamma f_0(1710)\rightarrow
\gamma 4\pi)$
, Crystal Barrel data\cite{Bar} on
${\bar p}p\rightarrow \pi^0 f_0(1500)\rightarrow \pi^0
(2\pi,{\bar K}K,2\eta,\eta'\eta,4\pi) $
 show $f_0(1500)$ decays into $4\pi$ dominantly through
sigma channel(S partial wave), and also
 hints it strongly couples to
$2\sigma$.

In this paper we take an effective lagrangian approach and
discuss the glueball decay into two-pion final states.
Particularly we will pay attention to the mixing between the
glueball and the sigma meson. The mixing angle is correlated
with the  decay width of the glueball into two pions. On the
other hand, by using the low energy theorem and QCD Sum Rule, we can
estimate the mixing angle. Then, some very useful
informations for those candidates can be obtained.

To begin with,  we consider, for simplicity, the linear sigma model for
two
flavors.
 As usual we introduce
the field
\begin{equation}
\Phi=\displaystyle{\sigma\frac{\tau^0}{2}+i{\vec \pi}\frac{\vec \tau}{2}},
\end{equation}
where $\tau^0$ is unity matrix and $\vec \tau$ the Pauli matrices with
normalization condition
  $Tr{\tau^a\tau^b}=2\delta^{ab}$.
Under a $SU_L(2)\otimes SU_R(2)$ chiral transformation, $\Phi$
transforms as
\begin{equation}
\Phi\rightarrow L^\dagger\Phi R .
\end{equation}
A renormalizable lagrangian of linear sigma model is given now by
\begin{eqnarray}
L_{\Phi-\Phi}=\displaystyle{Tr\{\partial_\mu\Phi^\dagger\partial^\mu\Phi\}
-\lambda[Tr\{\Phi^\dagger\Phi\}-\frac{f_\pi^2}{2}]^2},
\label{lal}
\end{eqnarray}
where $f_\pi$ is the vacuum expectation value of the
sigma field and $\lambda$ the self coupling
constant.

Let's now add the $O^{++}$  "pure" glueball state H
on the lagrangian.
Given that the
H is made of gluons and singlet under the chiral symmetry, there are
only two terms which give rise to the interaction among H,
the sigma and pions,
\begin{eqnarray}
L_{H-\Phi}=\displaystyle{g_1 f_\pi H[Tr\{\Phi^\dagger\Phi\}
-\frac{f_\pi^2}{2}]+g_2 H^2[Tr\{\Phi^\dagger\Phi\}
-\frac{f_\pi^2}{2}]},
\label{laH}
\end{eqnarray}
where two free parameters, $g_1$ and
$g_2$ are introduced
to describe the strength of couplings of one and two H to the
sigma or pions.

For the self interaction of glueballs, the lagrangian is given by
\begin{eqnarray}\label{self}
L_{H-H}=\displaystyle{\frac{1}{2}\partial_\mu H\partial^\mu H+
\frac{1}{2}m_H^2H^2+f_3H^3+f_4H^4},
\end{eqnarray}
  where $m_H^2$ is the mass of glueball, $f_3, f_4$ the self coupling constants.

After the chiral symmetry is broken by non-vanishing vacuum expectation
value of the sigma field, lagrangian in (\ref{laH}) generates not only
the interaction
for glueball decay, but also a mixing between the glueball
 H and the sigma field
$\sigma$. The mass matrix
for $(H, \sigma)$ is
\begin{equation}
m^2=
\left(
\begin{array}{cc}
m_H^2 & g_1f^2_\pi\\
g_1f^2_\pi & m^2_\sigma
\end{array}
\right ) ,
\end{equation}
where $m_\sigma^2=2\lambda f^2_\pi$.

To diagonalize the mass matrix,
we introduce an unitary matrix
\begin{equation}
V=
\left(
\begin{array}{cc}
cos\theta & sin\theta\\
-sin\theta & cos\theta
\end{array}
\right )
\sim\left(
\begin{array}{cc}
1 & \theta\\
-\theta & 1
\end{array}
\right ) .
\end{equation}
 Then the mass eigen-states are given by
\begin{equation}
\left(
\begin{array}{c}
H'\\
\sigma'
\end{array}
\right )
=
\left(
\begin{array}{cc}
1 & \theta\\
-\theta & 1
\end{array}
\right )
\left(
\begin{array}{c}
H\\
\sigma
\end{array}
\right ) ,
\label{mass}
\end{equation}
with $\theta= g_1 f_\pi^2 / (m_H^2 - m_\sigma^2)$.

Substituting (\ref{mass}) into (\ref{laH}) and (\ref{self}), we obtain the
lagrangian
 for $H'$ decay
\begin{equation}
L_{H'}=\displaystyle{H'\theta\{\frac{m_{H'}^2}{2 f_\pi}{\bf
\pi}^2+(\frac{m_{H'}^2+2m^2_{\sigma'}}
{2 f_\pi }- 2 g_2 f_\pi ){\sigma'}^2 +O(\theta^2)\}} .
\label{decay}
\end{equation}

In (\ref{decay}),
  the couplings of the glueball to sigma and the pions
 are proportional to the
mixing angle $\theta$. This could be understood that
the mixing between $H$ and $\sigma$
is of the same order as that for
 $H'$ to decay into light hadrons (see figure 1).
Furthermore,
under the assumption of vector meson dominance\cite{pisarski},
 $\rho$ meson can be introduced
by
 replacing $\partial^\mu\Phi$ in (\ref{lal})
by $D^\mu\Phi=\partial^\mu\Phi-ig(\rho^\mu\Phi-\Phi \rho^\mu)$. We found
that
the direct coupling of
 $H'$
and $\rho\rho$ vanishes at the tree level.

Identifying the glueball state
$H^\prime$ with the candidates, the phenomenologies of our
model can be
summarized as follows:

i) Glueball decay into two pions: The decay width of the
glueball into two pions is given by
\begin{equation}\label{width}
\Gamma ( H^\prime \rightarrow 2\pi )
= \frac{3\theta^2}{32 \pi}
(\frac{m_{H'}}{f_\pi})^2
\sqrt{m_{H^\prime}^2 -4 m_\pi^2}.
\end{equation}

Experimentally,  the data of $f_0(1370)$  are\cite{data}
\begin{equation}
\Gamma_{tot}=200-500MeV~~~~~~~~~~~~~
\displaystyle{\frac{\Gamma(\pi\pi)}{\Gamma_{tot}}=0.26\pm0.09}
\end{equation}
Therefore, $\Gamma(\pi\pi)\sim$ 100MeV. from (\ref{width}),
we obtain $\theta\sim$ 0.16.

For $f_0(1500)$, different groups
present controversial results. Let us use
 Crystal Barrel Group data\cite{Bar}.
$Br(f_0(1500)\rightarrow 2\pi):Br(f_0(1500)\rightarrow 4\pi)=
4.39\pm0.16:14.9\pm3.2$ and
 $\Gamma_{total}=120\pm20MeV$.
 Assuming that $Br( f_0(1500)\rightarrow 4\pi )$
is about $ 50\%$ one has that $\Gamma(2\pi)=17.5\pm 6\pm 3 MeV$.
Using this value, we obtain $\theta\sim 0.058$.

For $f_0(1710)$, the data are
\begin{equation}
\Gamma_{tot}=125MeV,~~~~~~~~~~~~~
\displaystyle{\frac{\Gamma(\pi\pi)}{\Gamma_{tot}}=0.039^{+0.002}_
{-0.024}}.
\end{equation}
Then,  $\theta\sim 0.025$.

ii)Glueball decay into four pions: the  decay of the glueball into
four pions is through the intermediate two sigma states.
In Figure 2, we
plot the ratio of glueball decay into  $2\sigma$ to $2\pi$ as
function
of $g_2$. This measurement can be used to determine the parameter
$g_2$.

For item i), an interesting application is to correlate it with
the estimate of the mixing angle $\theta$ by using QCD sum
rule.

The low energy theorem
states that \cite{shif}
\begin{equation}
i\int dx<|TO(x){\cal O}|>=-d_O<|O(0)|>,
\end{equation}
where $d_O$ is the mass dimension of operator $O$,
${\cal O}$
is the trace of the energy momentum tensor
$$\theta^\mu_\mu=\displaystyle{\frac{\beta(\alpha_s)}{4\alpha_s}
G^{\mu\nu}G_{\mu\nu} \approx
\frac{-b\alpha_s}{8\pi}G^{\mu\nu}G_{\mu\nu}},$$
($\beta(\alpha_s)$ is Gell-Mann-Low function, b=11 for pure Yang-Mills QCD)
which vanishes in classical
level.
To obtain the mixing angle $\theta$, let us firstly choose
operator $O$ to
be
 $O_\sigma=(\bar u u+\bar d d)/\sqrt{2}$.
Assuming that
the lowest-lying  $0^{++}$ $\bar qq$ state, i.e. $\sigma$ and
the lowest-lying $0^{++}$  glueball  saturate the
correlation function
we obtain
\begin{equation}
\displaystyle
{\langle|O_\sigma(0)|\sigma\rangle\frac{1}{m_\sigma^2}
\langle\sigma|{\cal
O}|\rangle +
\langle|O_\sigma(0)|H\rangle\frac{1}{m_H^2}
\langle H|{\cal
O}|\rangle
=-d_\sigma\langle|O_\sigma(0)|\rangle}.
\label{s}
\end{equation}
Similarly, taking now the operator $O$ to be ${\cal O}$, we have
\begin{equation}
\displaystyle
{\langle|{\cal O}(0)|H\rangle\frac{1}{m_H^2}\langle H|{\cal
O}|\rangle+
=-d_G\langle|{\cal O}(0)|\rangle}.
\label{g}
\end{equation}
In deriving (\ref{g}), we have also assumed that
the ground state of $0^{++}$  glueball saturates the
l.h.s of (\ref{g}). Defining $\langle|{\cal O}|H\rangle=f_H$,
$\langle|O_\sigma|\sigma\rangle=f_\sigma$,
we can roughly define the effective field of the "pure" glueball
${\cal H}$ as
\begin{equation}
\displaystyle{
{\cal H}=\frac{1}{f_H}{\cal O}-\frac{1}{f_H}\langle|{\cal
O}|\rangle.
}.
\end{equation}
Therefore,
\begin{equation}
\langle\sigma|{\cal O}|\rangle\approx \theta f_H .
\end{equation}
Similar,
\begin{equation}\label{o}
\langle H|O_\sigma|\rangle\approx -\theta f_\sigma .
\end{equation}

From eqs (\ref{s})-(\ref{o}), we get the mixing angle
\begin{equation}
\displaystyle{
\theta=-\frac{3m_\sigma^2m_H^2}{f_\sigma
f_H(m_H^2-m_\sigma^2)}\langle|O_\sigma|\rangle=
-\frac{3m_\sigma^2m_H}{2f_\sigma(m_H^2-m_\sigma^2)\sqrt{-\langle|{\cal
O}|\rangle}} -\langle| O_\sigma|\rangle
}.
\label{theta}
\end{equation}
Note that
the mixing angle $\theta$ is proportional to the quark's condense.
The parameter $f_\sigma$ in (\ref{theta}) can be estimated again by the QCD
sum rule. Following\cite{scalar}, we have
\begin{equation}
\begin{array}{ll}
f_\sigma^2=&\displaystyle{
\frac{6}{16\pi}M^4 \big (
1+\frac{13}{3}\frac{\alpha_s}{\pi}+
\frac{8\pi^2}{M^4}\langle|m\bar q q|\rangle
}\\
&\displaystyle{
+\frac{\pi^2}{3M^4}\langle|\frac{\alpha_s}{\pi}
G^{\mu\nu}G_{\mu\nu}|\rangle-
\frac{1408}{81}\frac{\pi^3\alpha_s}{M^6}\langle|\bar q
q|\rangle\big )e^{m_\sigma^2/M^2}},
\end{array}
\label{fs}
\end{equation}
where $M$ is a parameter with mass dimension in Borel
transformation.

For the quark and gluon condensate, we take
the standard values
\begin{equation}
\begin{array}{l}
\langle|m\bar q q|\rangle=-(0.1GeV)^4,\\
\langle|\bar q q|\rangle=\langle|\bar u u|\rangle=\langle|\bar
d d|\rangle=-(0.25GeV)^3,\\
\displaystyle{
\langle|\alpha_s G^{\mu\nu}G_{\mu\nu}|\rangle
=0.06\pm0.02GeV^4}.\\
\end{array}
\end{equation}
The mass prediction of sigma is in the region 700MeV-1000MeV,
where when M=500MeV, $m_\sigma=750$MeV and  M=1000MeV,
$m_\sigma=1000$MeV. This is consistent with the particle
data book. We also note that some recently experimental
fits give a lower sigma mass. However, N.A. T\"ornqvist gave a
sigma mass $m_\sigma$=860MeV by using the sigma
model\cite{torn}. Therefore, in our case QCD sum rule's
prediction is not bad. Let us take our discussion at
$m_\sigma$=860MeV. (\ref{theta}) gives
\begin{equation}
\begin{array}{ll}
\theta=0.10,~~~~~~~~~& m_H=1.37GeV,\\
\theta=0.086,~~~~~~~~~& m_H=1.5GeV,\\
\theta=0.068,~~~~~~~~~& m_H=1.71GeV
\end{array}
\label{va}
\end{equation}

In general QCD sum rule's calculation is
reliable within 30 percent uncertainty. Comparing the values in (\ref{va})
and
the mixing angles obtained by Eq.(\ref{width}) in i), we will be able to
exclude
$f_0(1710)$ as a glueball. It looks more like a $\bar ss$ meson.
Both $f_0(1500)$ and $f_0(1370)$ are
possible glueball states, but  we incline to $f_0(1500)$ is a
glueball, because there is still large uncertainty in its
experimental data. It is interesting to compare our result with
other approaches. In reference\cite{close1}, authors analyzed states
$f_0(1370)$, $f_0(1500)$ and $f_0(1710)$ by combining the various
experimental data and obtained a mixing pattern in terms of
pure glueball and other pure quark bound states  to be
\begin{eqnarray}
\begin{array}{llll}
              & f_{i1}^{G_s} & f_{i2}^{S} & f_{3i}^{(N)}\\
f_0(1710)     & 0.39 \pm 0.03 & 0.91\pm 0.02& 0.15\pm 0.02\\
f_0(1500)     & -0.65\pm 0.04& 0.33\pm 0.04& -0.70\pm 0.07\\
f_0(1370)     & -0.69\pm 0.07& 0.15\pm0.01&0.70\pm 0.07
\end{array}
\label{mix}
\end{eqnarray}
One can find our result by a simple approach is quite consistent with their
analysis. In a more recent paper\cite{new}, it was
pointed out that $f_0(1500)$ is a gluonium state and $f_0(1710)$ is a $\bar
ss$ state because  it is absent in $\bar pp$ annihilation. Meanwhile,
$f_0(1710)$'s $\bar ss$ dominance is consistent with $\gamma\gamma$ data
since only few decay branching ratios have been measured for this state.

Before concluding our paper, we make two remarks:

1) In the chiral limit which we work on in
this paper, the mixing of glueball with
sigma is a consequence of the spontaneous chiral symmetry breaking.
However, when including the explicit symmetry breaking effects by the quark
mass, especially
when extending the lagrangian for two flavors to three flavors, there would be
one term proportional to the explicit symmetry breaking which also generates
mixing of the glueball with the sigma meson as shown in Fig. 3;

2) Other $\bar qq$ scalar mesons are possibly included in our Lagrangian.
There will be  the mixing between the "pure" glueball with these
states.  If these new scalar mesons are $\bar uu +\bar dd$ mesons, the
decay of glueball into two pions can be via these states, therefore the
mixing angle between the glueball and sigma determined by (\ref{width})
will be reduced. On the other hand, in (\ref{s}) we also need insert
these states into the correlator, then mixing angle between the glueball and
sigma estimated by the low energy theorem  is also reduced. We cannot
say these two effects cancel each other completely, however, since we
assume that sigma is the lightest $\bar uu+\bar dd$ meson, the glueball's
decay inclines to via the $\sigma$ channel. The more interesting result
is that,  the glueball's decay
width can be reasonable  with a small mixing with $\bar qq$
meson. That hints the glueball is still probably "pure".
The mixing between the glueball and $\bar ss$ is not relevant to
the decay of glueball into pions, it can be included in SU(3) model.

In conclusion, we have studied the decay of the scalar glueball into
pions and its mixing
 with the sigma particle in effective theory. We fit our model
to the  experimental data on the decay of glueball
candidates, $f_0(1370)$,
$f_0(1500)$ and $f_0(1710)$ into two pions
and obtain  the corresponding  mixing angle  of glueball with
sigma.  By using  our estimation for the mixing angle  based on
low energy theorem and QCD sum rule, we conclude
that $f_0(1710)$ is not  a glueball and the most possible
glueball candidate is $f_0(1500)$.   We also show that
a glueball  decays into four pions through
2 $\sigma$ intermediate states. The $\rho$ can contribute to the
four-pion decay mode at one-loop level with branch ratio
\begin{equation}
\displaystyle {\frac{\Gamma(2\rho)}{\Gamma(2\sigma)}\sim 0.01}.
\end{equation}

{\bf ACKNOWLEDGMENTS}\\
 This
work was supported in part by the national natural science foundation of
China.

\newpage

{\bf Figure Captions}\\
Fig.1 (a) Illustration of glueball decay into $2\sigma$ or $2\pi$;
(b)Illustration of glueball mixing with $\sigma$ induced by the
quark condense.\\
Fig.2 Ratio of $f_0(1500)$ decay into $2\sigma$ to $2\pi$ as function
of $g_2$. In the numerical calculation, we take  $m_\sigma=500GeV$\\
Fig.3 Illustration of glueball mixing with $\sigma$ induced by the quark
masses.

\topmargin 0.1cm
\begin{figure*}[thb]
\epsfysize=6in
\epsffile{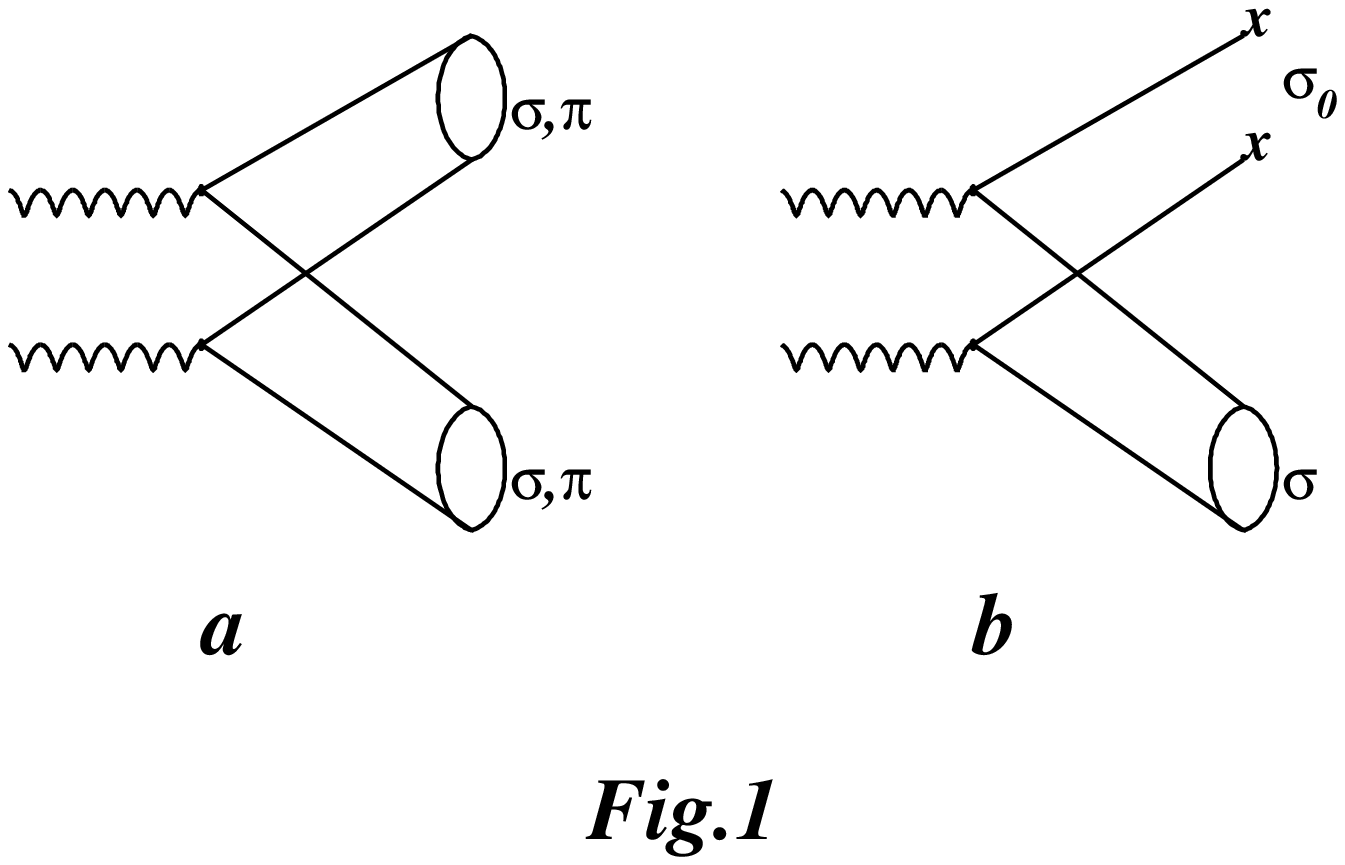}
\vskip 0.3cm
\end{figure*}

\topmargin 0.1cm
\begin{figure*}[thb]
\epsfysize=6in
\epsffile{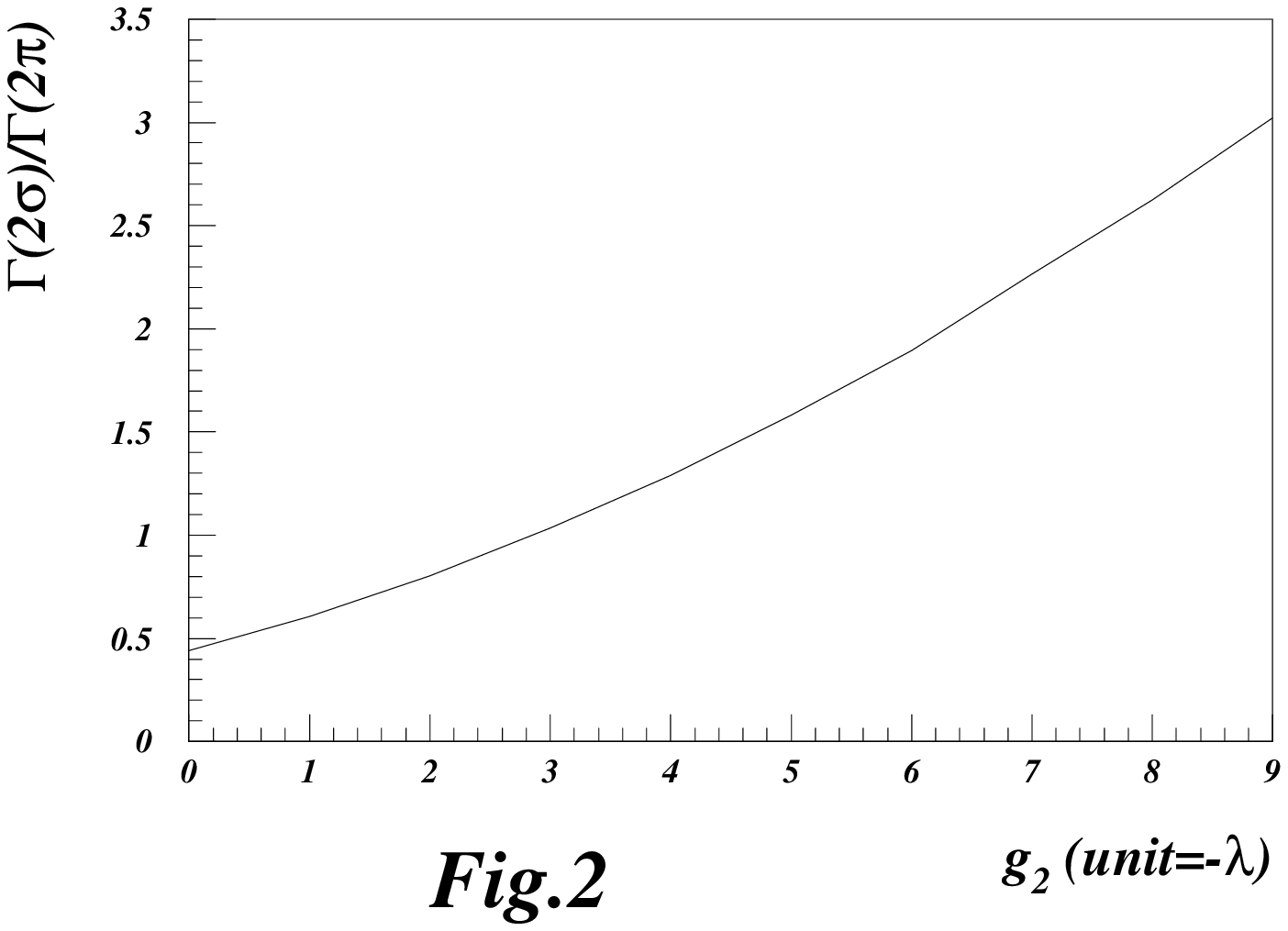}
\vskip 0.3cm
\end{figure*}

\topmargin 0.1cm
\begin{figure*}[thb]
\epsfysize=6in
\epsffile{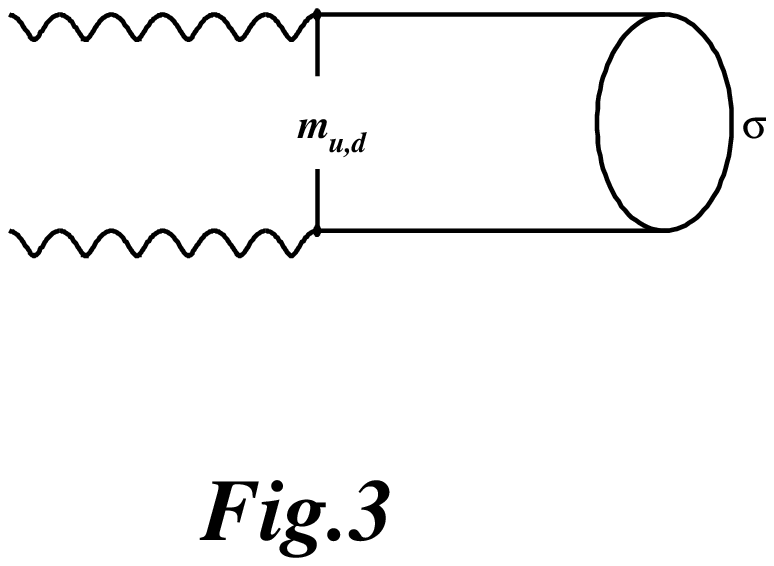}
\vskip 0.3cm
\end{figure*}

\end{document}